\documentclass[pra,aps,twocolumn,showpacs,floatfix,tightenlines]{revtex4}

\usepackage{amsmath}
\usepackage{epsfig}

\newcommand{\ket}[1]{\mbox{$|#1\rangle$}}
\newcommand{\bra}[1]{\mbox{$\langle #1|$}}

\newcommand{\ketbra}[2]{\mbox{$|#1\rangle\langle #2|$}}

\begin{document}

\title{Measuring Controlled-NOT and two-qubit gate operation}
\author{Andrew~G.~White, Alexei~Gilchrist, Geoffrey~J.~Pryde, Jeremy~L.~O'Brien, Michael~J.~Bremner, and Nathan~K.~Langford}
\affiliation{Centre for Quantum Computer Technology, University of Queensland, Brisbane, Queensland 4072, AUSTRALIA \\ andrew.white@uq.edu.au \phantom{andrew} www.quantinfo.org}

\begin{abstract}
Accurate characterisation of two-qubit gates will be critical for any realisation of quantum computation. We discuss a range of measurements aimed at characterising a two-qubit gate, specifically the \textsc{cnot} gate. These measurements are architecture-independent, and range from simple truth table measurements, to single figure measures such as the fringe visibility, parity, fidelity, and entanglement witnesses, through to whole-state and whole-gate measures achieved respectively via quantum state and process tomography. In doing so, we examine critical differences between classical and quantum gate operation.
\end{abstract}
\maketitle

\section{Introduction}
A notable aspect of current quantum information science is the quest to realise a scaleable quantum computing architecture. In the most common paradigm, the two required elements  are a single-qubit arbitrary rotation gate and a two-qubit maximally entangling  gate, e.g. a controlled-\textsc{not} (\textsc{cnot}) gate \cite{MikeIke}, although more generally any non-maximally entangling gate can be used \cite{Bry,Bremner}. There are many two-level quantum systems suitable for encoding qubits and realising single qubit rotations --- a non-exhaustive list includes \cite{Overview}: spin $\frac{1}{2}$ systems (e.g. electronic and nuclear spin); simple harmonic oscillators (e.g. atomic or molecular energy levels; phonon energy levels); superconducting systems (e.g. charge, phase and flux); and optical systems (e.g. photon polarisation). Whatever the choice of qubit system, it is necessary to demonstrate an entangling gate: consequently, a wide variety of experimental architectures are currently under consideration for realising \textsc{cnot} and other two-qubit gates \cite{Overview}.

Common to all these architectures is the need for an accurate characterisation of the systems that generate the universal gate set. A quantum gate is a unitary operation that, by definition, maps pure states to pure states. In the standard model a two-qubit quantum gate transforms states of two qubits to states of two qubits. In principle, a two-qubit quantum gate could output superposition states, entangled states and could also disentangle states --- depending on the nature of the gate and the input states. The task of determining if a prototype gate is working is not a simple matter. The prototype may decohere the qubits, not generate the correct amount of entanglement, or  introduce phase errors --- there are a myriad of potential experimental faults that might occur. In this paper we will consider a number of experiments that are focussed on detecting signatures of a \textsc{cnot} gate. In sections \ref{sec:simple} and \ref{sec:state}, we will consider a range of experiments that examine the nature of output states of a prototype \textsc{cnot} with a view to detecting properties that are characteristic of output states of a \textsc{cnot} operation. The discussion is arranged in order of increasing difficulty. Specifically, we will look at the logical operations performed by a \textsc{cnot}, the visibility of the output states and a number of experiments focussed on identifying entanglement generation between two qubits --- arguably the most important property of \textsc{cnot} gates. In section \ref{sec:state} we will discuss quantum state tomography and methods for detecting signatures of errors in a prototype gate by analysis of the structure of the output density matrices. Finally in section \ref{sec:process} we will discuss quantum process tomography, a more complete methodology for the characterisation of a quantum operation. We note that a similar set of arguments can be applied to derive appropriate measures for any two-qubit entangling gate. 

\section{Simple measures}
\label{sec:simple}

\subsection*{Truth table}

\begin{table}
 \begin{center}
 \begin{tabular}{|c|c|} \hline
 $($CT$)_{in}$ & $($CT$)_{out}$ \\ \hline
 0 \ 0 &  0 \ 0 \\
 0 \ 1 &  0 \ 1 \\
 1 \ 0 &  1 \ 1 \\
 1 \ 1 &  1 \ 0 \\ \hline
 \end{tabular}
 \end{center}
\caption{Input-output table for an ideal classical \textsc{cnot} (\textsc{xor}) gate. CT $\equiv$ respective values of Control \& Target bits.}
 \label{tab:cCNOT}
\end{table}

The exclusive-\textsc{or} (\textsc{xor}) gate is the classical two-bit \textsc{cnot} gate, where a control bit flips the state of a target bit. As shown in Table \ref{tab:cCNOT}, there are only four possible input, and therefore output, states. The simplest characterisation of an unknown two-bit gate is a straightforward matter of entering each possible input and measuring the output state --- if it conforms to Table \ref{tab:cCNOT}, it is at least an \textsc{xor} gate. A more complete characterisation, which allows measurement of error probabilities, is to measure the probability of each of the four possible output states for each of the four inputs, yielding a truth table as shown in Table \ref{tab:cCNOT2}. One measure of the overlap between a measured truth table, $M_{exp}$ and the ideal truth table, $M_{0}$, is given by the average of the the logical basis fidelities, the {\em inquisition}, $\mathcal{I} = \mathrm{Tr}(M_{exp} M_{0}^{T})/4$. Recent experiments have achieved inquisitions of 73\% \cite{ionCNOT} and 84\% \cite{photonCNOT}.

In the quantum case, the inputs are qubits, which can exist in an arbitrary complex superposition of a classical bit, e.g. $| \psi \rangle^{c}_{in} = \alpha |0 \rangle + \beta |1 \rangle$, where $|\alpha|^2+|\beta|^2=1$. Thus the gate has infinitely many possible inputs, and characterisation is not a simple matter of exhausting all possible inputs. However the difference to the classical gate is deeper than this. As Table \ref{tab:qCNOT} shows, the output states of a quantum CNOT gate can be entangled --- i.e. states with correlations that may not necessarily be replicated by classical models of physics. 

\begin{table}
 \begin{center}
 \begin{tabular}{|c|cccc|} \hline
    & $($00$)_{out}$ &  $($01$)_{out}$ &  $($10$)_{out}$ &  $($11$)_{out}$ \\ \hline
$($00$)_{in}$   &  1 & 0 & 0 & 0 \\
$($01$)_{in}$   &  0 & 1 & 0 & 0 \\
$($10$)_{in}$   & 0 & 0 & 0 & 1 \\
$($11$)_{in}$   & 0 & 0 & 1 & 0 \\ \hline
 \end{tabular}
 \end{center}
\caption{Truth table for an ideal classical \textsc{cnot} (\textsc{xor}) gate. The numbers indicate the probability of achieving the selected output state for a given input state.}
\label{tab:cCNOT2}
\end{table}

\begin{table}
\begin{center}
\begin{tabular}{|l|c|c|} \hline
Input label   &  $|C T\rangle_{in}$ & $|C T\rangle_{out}$ \\ \hline
                      &  $|00\rangle$ &  $|00\rangle$\\
Logical         &  $|01\rangle$ &  $|01\rangle$\\
                      &  $|10\rangle$ &  $|11\rangle$\\
                      &  $|11\rangle$ &  $|10\rangle$\\ \hline
Control- &  $( \alpha |0\rangle+ \beta |1\rangle ) |0\rangle$ & $\alpha |00\rangle+ \beta |11\rangle$ \\
superposition &  $( \alpha |0\rangle+ \beta |1\rangle ) |1\rangle$ & $\alpha |01\rangle+ \beta |10\rangle$ \\ \hline
Control- \& Target- & $( \alpha |0\rangle+ \beta |1\rangle )$ & $\phantom{+} \alpha \gamma |00\rangle+ \alpha \delta |01\rangle$ \\
superposition & \phantom{@} $\cdot ( \gamma |0\rangle+ \delta |1\rangle )$  & $+ \beta \delta |10\rangle + \beta \gamma |11\rangle$ \\ \hline
\end{tabular}
\end{center}
\caption{Input-output table for a quantum \textsc{cnot} gate: $|C T\rangle_{in,out} \equiv$ respective value of  input and output quantum states. With the {\em logical} inputs the table is as for a classical \textsc{cnot}. With {\em control-superposition} inputs, the outputs vary between separable and maximally-entangled. N.B. $|\alpha|^2$+$|\beta|^2$=$1$, $|\gamma|^2$+$|\delta|^2$=$1$. Note that since the \textsc{cnot} gate is reversible, the input and outputs can be swapped and the table is still correct --- a  \textsc{cnot} can coherently entangle or disentangle.}
 \label{tab:qCNOT}
\end{table}

\subsection*{Visibility and Parity}

A signature of quantum gate operation is the generation of entangled output states -- thus it is necessary to identify, and preferably quantify, these outputs. A beginning point is to measure a series of correlations, or {\em coincidence probabilities}, between the control and target arms, with the aim of identifying uniquely  quantum correlations. Consider, for example, the general case of a control-superposition input (row 5 of Table \ref{tab:qCNOT}), where we rewrite the states as $|\psi\rangle_{C}$=$(|0\rangle+ \varepsilon |1\rangle)/\sqrt{1+|\varepsilon|^2}$ and $|\psi\rangle_{T}$=$| 0 \rangle$. A \textsc{cnot} gate outputs the entangled state $|\psi\rangle_{out}$=$(|00\rangle+ \varepsilon |11\rangle)/\sqrt{1+|\varepsilon|^2}$  (where $\varepsilon$ is a complex number and the state is maximally entangled if $|\varepsilon|$=$1$).  Note that, for a target initially in the logical 0 state, the output state never contains terms with odd parity, only even parity terms, i.e. the control and target are always either both 0, or both 1. Thus the correlation between the arms can be quantified by constructing a coincidence fringe {\em visibility} in the logical basis \cite{Comb}, e.g.,
\begin{equation}
V_{L} =  \frac{P_{\mathrm{even}}-P_{\mathrm{odd}}}{P_{\mathrm{even}}+P_{\mathrm{odd}}} = \frac{P_{00}-P_{01}}{P_{00}+P_{01}} = \frac{(1+|\varepsilon|^2)^{-1}}{(1+|\varepsilon|^2)^{-1}}  = 1,
\label{eq:VL}
\end{equation}
where $P_{ij}=\mathrm{Tr}(\hat{\rho}|i j\rangle \langle i j|)=|\langle i j| \psi \rangle_{out}|^{2}$ is the coincidence probability of seeing the state $|i j\rangle$ where $i,j = 0$ or $1$. This is a simple measure of a putative \textsc{cnot} gate: what is the logical basis visibility? If it is not unity, the gate does not always act as a \textsc{cnot} gate.

To detect entanglement it is necessary to input {\em and} measure superposition states. We will consider equal-weight superpositions, since they are furtherest from the logical bases \cite{CircleNote}. Analysing with arbitrary superposition states, $|\theta \rangle_{i} = \cos \theta_{i} |0\rangle_{i} + e^{i \phi_{i}} \sin \theta_{i} |1\rangle_{i}$, where $i=C$ or $T$, the coincidence probability becomes \cite{Nonmax},
\begin{eqnarray}
\rm{P}_{\theta_{C} \theta_{T}}
& = &
|\langle \theta_{C}| \langle \theta_{T}| \psi \rangle_{out}|^{2} \\
& = & \frac{|\cos \theta_{C} \cos \theta_{T} + \varepsilon^{*} e^{-i (\phi_{C}+\phi_{T})} \sin \theta_{C}
\sin \theta_{T}|^{2}}{(1+|\varepsilon|^{2})} \nonumber .
\label{eq:coinc 1}
\end{eqnarray}
Equal-weight analysis occurs when $\theta_{C}$=$\theta_{T}$=$\frac{\pi}{4}$, and we define the {\em equal-weight visibility} to be,
\begin{eqnarray}
V_{E}(\phi_{C},\phi_{T})  & = & \frac{P_{\phi_{C},\phi_{T}}^{\frac{\pi}{4}} - P_{\phi_{C},\phi_{T}+\pi}^{\frac{\pi}{4}}}{P_{\phi_{C},\phi_{T}}^{\frac{\pi}{4}} + P_{\phi_{C},\phi_{T}+\pi}^{\frac{\pi}{4}}} \nonumber \\
 & = & \frac{2 |\varepsilon|}{1+ |\varepsilon|^{2}} \cos(\phi_{C}+\phi_{T}+\xi),
\label{eq:VE}
\end{eqnarray}
where $\xi$=$\mathrm{arg}(\varepsilon)$. This is an indicator of quantum \textsc{cnot} gate operation: in the ideal case if the control state is an equal-weight superposition, $|\varepsilon|=1$, then $|V_{E}|=1$ for appropriate $\phi_{C},\phi_{T}$. In the special case where the control and target analysers have the same phase, $\phi_{C}$=$\phi_{T}$=$\phi$, the visibility becomes,
\begin{equation}
V_{E}(\phi)  = \frac{2 |\varepsilon|}{1+ |\varepsilon|^{2}} \cos(2 \phi+\xi).
\label{eq:VE2}
\end{equation}
A similar measure is the {\em parity} \cite{4ion,ionCNOT}, which for two qubits is,
\begin{eqnarray}
\Pi(\phi) & = & P_{\phi,\phi}^{\frac{\pi}{4}} + P_{\phi + \pi,\phi + \pi}^{\frac{\pi}{4}} - P_{\phi,\phi + \pi}^{\frac{\pi}{4}} - P_{\phi + \pi,\phi}^{\frac{\pi}{4}}.
\label{eq:parity1}
\end{eqnarray}
For our example this reduces to $\Pi(\phi)  = V_{E}(\phi)$, i.e., for two qubits, the parity is equivalent to the equal-weight visibility. As the phase $\phi$ of the equal-weight analysers is varied, $V_{E}$ and $\Pi$ oscillate with an amplitude which is twice the magnitude of the extreme off-diagonal element of the density matrix, $| 0 ... 0 \rangle \langle 1 ... 1 |$ \cite{4ion}, and a frequency of $2 \phi$.

The latter behaviour has been used as an indicator of quantum operation \cite{ionCNOT}, but care must be exercised using this as a measure in the two-qubit case as separable output states can also give $2 \phi$ fringes (e.g. for $|\psi_{in} \rangle = | 0+1 \rangle | 0-1 \rangle/2$ the parity of the output is $\Pi(\phi) \propto \cos^{2} \phi = (1+\cos 2 \phi)/2$ ). The following argument highlights the need for caution. Consider a semi-classical \textsc{cnot} gate, which measures the control qubit and applies a bit flip ($X$) if the result is 1. This is described by the completely positive (CP) map,
\begin{equation}\label{semi-cnot}
\begin{split}
\mathcal{E}(\hat{\rho}) =& (K_{0}\otimes I)\hat{\rho}(K_{0}\otimes I) + (K_{1}\otimes X)\hat{\rho}(K_{1}\otimes X),
\end{split}
\end{equation}
where $K_{0}$=$\ketbra{0}{0}$ and $K_{1}$=$\ketbra{1}{1}$, $\hat{\rho}$ is the density matrix of the input state, and the one-qubit identity operator, $I$, and the Pauli spin operators, $X,Y,Z$ are given by,
\begin{eqnarray}
\begin{array}{cc}
I \equiv \hat{\sigma}_{0} = \left[\begin{array}{cc} 1 & 0 \\ 0 & 1 \end{array}\right] &
X \equiv \hat{\sigma}_{1} = \left[\begin{array}{cc} 0 & 1 \\ 1 & 0 \end{array}\right] \nonumber \\ \nonumber \\
Y \equiv \hat{\sigma}_{2} = \left[\begin{array}{cc} 0 & -i \\ i & 0 \end{array}\right] &
Z \equiv \hat{\sigma}_{3} = \left[\begin{array}{cc} 1 & 0 \\ 0 & -1\end{array}\right].
\label{eq:su2}
\end{array}
\end{eqnarray}

Gate (\ref{semi-cnot}) reproduces the action of the \textsc{cnot} gate in the computational basis, and all superpositions of target qubit. However, it has a parity of zero for the control-superposition inputs (lines 5 and 6 of Table \ref{tab:qCNOT}), since it destroys the superposition in the control qubit. It is easy to invert the orientation of the gate by applying Hadamard gates ($\mathcal{H}$) before and after, i.e.,
\begin{equation}\label{semi-cnot-inv}
\begin{split}
\mathcal{E}(\hat{\rho}) = & (I\otimes \mathcal{H} K_{0} \mathcal{H})\hat{\rho}(I\otimes \mathcal{H} K_{0} \mathcal{H}) +\\
& (Z \otimes \mathcal{H} K_{1} \mathcal{H})\hat{\rho}(Z\otimes \mathcal{H} K_{1} \mathcal{H}).
\end{split}
\end{equation}
The parity of this strictly semi-classical gate has the same number of fringes as an ideal \textsc{cnot} gate --- so parity alone cannot be used as an indicator of quantum operation. Fortunately, this gate can be distinguished from a true \textsc{cnot} by the results in the computational basis. Therefore, measurement of both the computational basis and the parity are required \cite{ionCNOT}.

\subsection*{Bell-state Fidelity}

Given that entanglement is a signature of quantum gate operation, an unambiguous indicator of its presence is desirable. One such indicator is the {\em fidelity} of any of the two-qubit Bell states ($|\phi^{\pm}\rangle=(|00 \rangle \pm |11\rangle)/\sqrt{2}$ and $|\psi^{\pm} \rangle = (|01 \rangle \pm |10 \rangle)/\sqrt{2}$) with the measured state, $\hat{\rho}$, i.e.
\begin{eqnarray}
F_{\phi^{\pm}} & = & \mathrm{Tr}(\hat{\rho} \ |\phi^{\pm} \rangle \langle \phi^{\pm}|) \\ \nonumber 
F_{\psi^{\pm}} & = & \mathrm{Tr}(\hat{\rho} \ |\psi^{\pm} \rangle \langle \psi^{\pm}|).
\label{eq:fid1}
\end{eqnarray}
For any separable state, $F_{\phi^{\pm},\psi^{\pm}} \leq \frac{1}{2}$: if $F_{\phi^{\pm},\psi^{\pm}} > \frac{1}{2}$, the state is entangled (for instance, see lemma 1 in \cite{Terhal00a}). It is not necessary to measure the full density matrix, $\hat{\rho}$ (which requires 16 coincidence probabilities, see next section) to determine the Bell fidelities --- they can be calculated from just six coincidence probabilities,
\begin{eqnarray}
F_{\phi^{\pm}} & = & \frac{P_{HH}+P_{VV} \pm P_{DD} \pm P_{AA} \mp P_{RR} \mp P_{LL}}{2} \label{eq:fid2} \\
F_{\psi^{\pm}} & = & \frac{1 \mp 1 -P_{HH} - P_{VV} \pm P_{DD} \pm P_{AA} \pm P_{RR} \pm P_{LL}}{2} \nonumber
\end{eqnarray}
where $P_{ii}$ is the coincidence probability in the $i$ basis and we adopt optical-polarisation nomenclature as shorthand and define: the logical basis as {\em horizontal} and {\em vertical}, $|H\rangle$$\equiv$$|0\rangle$ \& $|V\rangle$$\equiv$$|1\rangle$; the equal-weight real superpositions as {\em diagonal} and {\em anti-diagonal}, $|D\rangle$=$(|0\rangle + |1\rangle)/\sqrt{2}$ and $|A\rangle$=$(|0\rangle - |1\rangle)/\sqrt{2}$; and the equal-weight complex superpositions as {\em right} and {\em left}, $|R\rangle$=$(|0\rangle + i |1\rangle)/\sqrt{2}$ and $|L\rangle$=$(|0\rangle - i |1\rangle)/\sqrt{2}$. Even in the presence of experimental uncertainty, the Bell state fidelity is a reliable and experimentally robust indicator of entanglement (see next section). A caveat regarding the  Bell Fidelities is that they are only sensitive to the state they represent and will miss some entangled states. For instance, the maximally entangled states $(\mathcal{H} \otimes I)\ket{\phi^\pm}$ or $(\mathcal{H} \otimes I)\ket{\psi^\pm}$, always result in $F_{\phi^{\pm},\psi^{\pm}}  \le  \frac{1}{2}$.

\subsection*{Entanglement Witness}

More generally, the presence of entanglement can be flagged by an \emph{entanglement witness} \cite{96horodecki1,00terhal319} --- in the usual formulation, an observable $W_\phi$ such that $\langle W_\phi \rangle <0$ for some entangled states $\ket{\phi}$ and $\langle W_\phi \rangle \ge 0$ for all unentangled states. For every entangled state a suitable entanglement witness exists.  An \emph{optimal} entanglement witness for a class of states is one that will detect the entanglement in more states in that class than any other witness.

Since we have \emph{a priori} knowledge of the entangled state that should be produced from the \textsc{cnot}, we will concentrate on detecting the entanglement in Werner states, i.e.,
\begin{equation}
  \label{eq:depolarised}
  \frac{1}{4} (1-p) I \otimes I + p \ketbra{\phi}{\phi}
\end{equation}
where $p \in [0,1]$, $I \otimes I$ is the maximally mixed state --- to model the effect of decoherence --- and $\ketbra{\phi}{\phi}$ is the maximally entangled state expected from an ideal \textsc{cnot}: this state models the effect of a decohering channel. An optimal entanglement witness for this class of states can be simply constructed \cite{02guhne062305} by finding the eigenvector $\ket{w}$ corresponding to the minimum eigenvalue of $\ketbra{\phi}{\phi}^{T_2}$, where $T_2$ is the partial transpose operation on the second qubit \cite{PTfoot}. The witness is then constructed from $W_\phi=\ketbra{w}{w}^{T_2}$.  For example, to detect $\ket{\phi^{+}}$ the entanglement witness is,
\begin{equation}
  \label{eq:Wphip}
  W_{\phi^{+}} =\frac{1}{2} 
\begin{pmatrix}
0&0&0&-1\\
0&1&0&0\\
0&0&1&0\\
-1&0&0&0
\end{pmatrix}.
\end{equation}
In order to measure this witness we want to decompose it into a sum of local measurements and minimise the number of measurements necessary. It is clear that for two qubits we can decompose the witness operator into tensor products of Pauli matrices --- hence only three detector settings are necessary. (In \cite{02guhne062305} it was shown that three settings is the minimum that can be achieved.) The following is the decomposition of the witnesses for the four Bell states:
\begin{align}
  W_{\phi^{\pm}} &= \frac{1}{4}(I\otimes I \mp X\otimes X \pm Y\otimes Y - Z\otimes Z)\\ \nonumber
  W_{\psi^{\pm}} &= \frac{1}{4}(I\otimes I \mp X\otimes X \mp Y\otimes Y + Z\otimes Z),
\end{align}
These equations show that, as with the Bell-state fidelities, it is not necessary to fully measure the density matrix to measure the witnesses --- a set of coincidence probabilities suffices. (This is particularly advantageous for experimental architectures where state tomography is difficult.) For example by using $\mathrm{Tr}\{ \varrho X \}=P_{D}-P_{A}$ where $\varrho$ is a single qubit density matrix, one such decomposition is:
\begin{eqnarray} 
  \langle W_{\phi^{\pm}} \rangle & = & \frac{P_{HV}+P_{VH}\mp P_{DD}\mp P_{AA}\pm P_{RR}\pm P_{LL} }{2} \nonumber \\ 
 & \phantom{=}& \label{eq:W2} \\ 
  \langle W_{\psi^{\pm}} \rangle & = & \frac{P_{HH}+P_{VV}\mp P_{DD}\mp P_{AA}\pm P_{RL}\pm P_{LR} }{2} \nonumber 
\end{eqnarray}
Entanglement witnesses can be recast so that $\langle W \rangle$$<$$r$ ($>$$r$), for a set of non-separable  states, where $r \in \Re$; and $\langle W \rangle$$\geq$$r$ ($\leq$$r$), for all other states (including the separable states). In fact, the Bell-state projectors are such a case of an entanglement witness (see Appendix \ref{sec:ew}), where,
\begin{equation}
\langle W_{\phi^{\pm},\psi^{\pm}} \rangle = \frac{1}{2}-F_{\phi^{\pm},\psi^{\pm}}.
\end{equation}
In general, {\em any} projective measurement that projects onto an entangled state can be used as an entanglement witness. As far as is known, all entanglement witnesses share the problem highlighted for the Bell-state fidelity --- due to their specificity, they will miss some entangled states.

\subsection*{Tangle and Bell's inequality}

The {\em tangle}, $T$, is a measure that detects any degree of entanglement, for all two-qubit states, be they pure or mixed \cite{Tangle1,Tangle2}. It is quantitative, varying between $T$=0 for a unentangled state to $T$=1 for a maximally entangled state. The Tangle is not normally considered a simple measure, as it requires measurement of the output two-qubit density matrix, $\hat{\rho}$ (see next section). However, for the special case of a perfectly coherent \textsc{cnot} gate --- admittedly unlikely in practice --- the tangle is simply given by $T=V_{E}(n \frac{\pi}{2})^{2}$ for $n$=$0,1,2...$

For pure output states the tangle is related to the Bell inequality \cite{Bell,CHSH}, $B$, by, $B=2 \sqrt{1+T^2}$ \cite{Munro}. For states that can be described by classical local hidden variable models, $B \leq 2$, whereas for a certain range of quantum states, including maximally-entangled states, $B>2$ ($B=2 \sqrt{2}$ for a maximally-entangled state). A violation of Bell's inequality is a very strong statement as it rules out a wide class of possible ``classical'' models, however to be conclusive in this regard, it requires high-efficiency space-like  separated measurements --- a feat not yet achieved in any experimental architecture.

\section{State measurements}
\label{sec:state}

What if a gate has a logical visibility less than unity? This may occur due to a variety of reasons, two possible extremes being that the gate operates non-deterministically (i.e only some of the time) as a \textsc{cnot} gate, or that the gate operates deterministically, but {\em not} as a \textsc{cnot} gate. Additional measurements are required to distinguish between these alternatives.

The measurement of the state of systems of $n$ qubits is a solved problem, dealt with in the comprehensive paper of James {\em et al.} \cite{James}. Here we summarise their results relevant to two-qubit gates. As in the last section, we use optical nomenclature for brevity, but this discussion applies equally to all qubits, independent of physical architecture.

The density matrix of a single qubit (or ensemble of identically prepared single qubits) is given by,
\begin{equation}
\hat{\rho} = \frac{1}{2} \sum_{i=0}^{3} \frac{S_{i}}{S_{0}} \hat{\sigma}_{i}
\label{rho1}
\end{equation}
where $\hat{\sigma}_{0}$ is the identity operator, and $\hat{\sigma}_{1,2,3}$ are the Pauli spin operators. The single-qubit {\em Stokes parameters} are given by,
\begin{equation}
S_{i} = 2 P_{i} - 1
\end{equation}
where $P_{i}$ are the 4 measurement probabilities represented by the four projectors $\hat{\mu}_{0}= |H \rangle \langle H| + |V \rangle \langle V|$, $\hat{\mu}_{1}= |H \rangle \langle H|$, $\hat{\mu}_{2}= |D \rangle \langle D|$, $\hat{\mu}_{3}= |R \rangle \langle R|$. (Note that this set is not unique, and other, more convenient sets may be used depending on architecture. For example, polarisation qubits are typically measured using $\hat{\mu}_{0}= |H \rangle \langle H|$, $\hat{\mu}_{1}= |V \rangle \langle V|$, $\hat{\mu}_{2}= |D \rangle \langle D|$, $\hat{\mu}_{3}= |R \rangle \langle R|$ \cite{Nonmax}).

Similarly, for two-qubits (or identical ensemble of same), the density matrix is given by,
\begin{equation}
\hat{\rho} = \frac{1}{2^{2}} \sum_{i_{1},i_{2}=0}^{3} \frac{S_{i_{1},i_{2}}}{S_{0,0}} \hat{\sigma}_{i_{1}} \otimes \hat{\sigma}_{i_{2}}
\label{rho2}
\end{equation}
where the two-qubit Stokes parameters are given by,
\begin{eqnarray}
S_{i_{1},i_{2}} & = & \sum_{j_{1},j_{2}}^{3} (\Upsilon^{-1})_{i_{1},j_{1}} (\Upsilon^{-1})_{i_{2},j_{2}} P_{i_{1},i_{2}}  \\
\mathrm{where} \ \Upsilon & = &
\begin{pmatrix}
      1 & 0 & 0 & 0\\
    -1  &  2 & 0 & 0 \\
    -1  &  0  & 2 & 0\\
    -1  &  0 & 0 & 2
\end{pmatrix},  \nonumber
\end{eqnarray}
and $P_{i_{1},i_{2}}$ are the 16 coincidence measurement probabilities represented by the projectors $\hat{\mu}_{i} \otimes \hat{\mu}_{j}$ ($i,j=0,1,2,3$).

Equations \ref{rho1} and \ref{rho2} are not applicable in the presence of experimental uncertainties --- such as those introduced by count statistics or analyser uncertainties --- if used in such a case they can lead to non-physical density matrices that violate properties such as positivity. Fortunately, this can be avoided using a more involved formulation involving maximum likelihood estimation  \cite{James} --- but one still based on the above probability measurements.

Having measured a density matrix, it is instructive to consider what physical information can be extracted from it. The most obvious attribute is the degree of order, or entropy. For a pure state, the degree of order is maximal and the entropy is  therefore zero. Entropy measures used to date include the {\em von Neumann} and {\em linear} entropies, in the latter case, the entropy is scaled between $S_{L}=0$ (for a pure state) and $S_{L}=1$ (for a fully mixed state) \cite{Hilbert}. For perfectly coherent \textsc{CNOT} operation, if the input states are pure then all output states will be pure and the measured entropies will be zero.

Single qubit density matrices can be uniquely decomposed into a maximally mixed component ($S_{L}=1$) and a completely pure component ($S_{L}=0$), e.g. polarised light can be described as a specific combination of unpolarised and purely polarised light. It is tempting to decompose two-qubit density matrices in a similar fashion, however in this case no unique decomposition exists --- infinitely many possible combinations are possible --- and any physical conclusions based on such a procedure should be treated with caution. Specifically, it is not possible to uniquely characterise the action of any two-qubit gate in this manner.

As we have seen, the other physically significant attribute of two-qubit systems is the degree of entanglement. There is an active literature on measures for quantifying entanglement: here we consider only the {\em tangle}, which varies between $T=0$ for all separable states to $T=1$ for all maximally-entangled states. For a density matrix in the logical basis representation, $\rho$, the spin-flipped matrix is defined to be \cite{Tangle2},
\begin{equation}
\tilde{\rho} = (Y \otimes Y) \hat{\rho}^{*} (Y \otimes Y),
\end{equation}
where $\hat{\rho}^{*}$ is the complex conjugate of $\hat{\rho}$. $\{ \lambda_{1}, \lambda_{2}, \lambda_{3}, \lambda_{4} \}$ are defined to be the eigenvalues of $\hat{\rho} \tilde{\rho}$ in decreasing order, and the tangle is 
\begin{equation}
T = (\mathrm{max} \{ \lambda_{1} - \lambda_{2} - \lambda_{3} - \lambda_{4}, 0 \})^{2}.
\end{equation}
Amongst other properties, the tangle is a conserved measure of the amount of entanglement that can be shared between multiple qubits, and the tangle is well-defined for two-qubit mixed states. James {\em et al.} \cite{James} have modelled the effect of experimental uncertainties on a number of quantities derived from the density matrix, including the entropy and tangle, making them suitable quantities to use in characterising experimental systems.
  
\begin{figure}
\center{\epsfig{figure=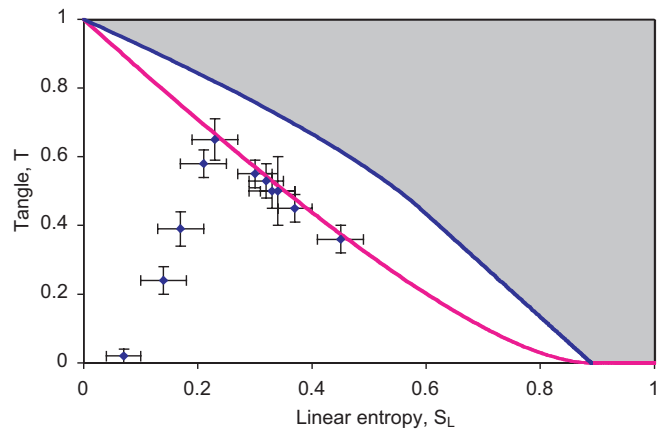,width=\columnwidth}}
\caption{{\em Tangle-entropy plane}. The data points are measured values for a range of two-qubit states (in this case, polarisation-entangled photons). The upper line indicates the maximally-entangled mixed states \cite{MEMS}, the lower line indicates Werner states (equation \ref{eq:depolarised}).}
 \label{fig:TS}
\end{figure}  
  
The salient attributes of a two-qubit state can be characterised by its location on the {\em tangle-entropy plane} \cite{Munro,Hilbert}. The output states of an ideal \textsc{cnot} gate lie on the ordinate axis; non-ideal operation introduces decoherence, and the measured output states move from the axis into the plane. There is a limit to the amount of entanglement possible for a given amount of decoherence, states that have the maximum amount are known as {\em maximally-entangled mixed states} (MEMS) and form a boundary on the tangle-entropy plane \cite{MEMS,MEMS2}.

Figure \ref{fig:TS} shows a range of two-qubit states --- in this case, polarisation-entangled photons --- of varying entropy and tangle obtained in our laboratory, using an experimental system essentially identical to that described in \cite{Nonmax,Hilbert}. Each data point comes from an experimentally derived density matrix, constructed from sixteen measurements as described above. Table {\ref{tab:expW}} lists the Bell-state fidelities for these matrices --- both those obtained with six measurements (equation \ref{eq:fid2}) and those obtained via the density matrix \cite {James}. Despite the presence of experimental uncertainties, the fidelity reliably witnesses the entanglement in every case. Not unsurprisingly, the six-measurement fidelity has less uncertainty than the sixteen-measurement --- in the latter case, more parameters are measured than are necessary to find the Bell-state fidelity, introducing additional uncertainty. With respect to gate characterisation, this table underscores the point made earlier that that the Bell-state Fidelities --- indeed, entanglement witnesses of any kind --- are not metrics for entanglement, i.e. they do not provide a {\em measure} of how entangled a state is, they merely provide an {\em indicator} that the state is entangled. 

 \begin{table}
 \begin{center}
 \begin{tabular}{|c|c|c|c|} \hline
Entropy & Tangle & Fidelity   & Fidelity  \\ 
$S_{L}$ & $T$      & 6 measurements &  16 measurements \\ \hline                                                                                                  
$0.07(3)$ & $0.02(2)$ & $F_{\phi^{+}}=0.509(12)$ & $F_{\phi^{+}}=0.54(9)$ \\ 
$0.14(4)$ & $0.24(4)$ & $F_{\phi^{+}}=0.716(13)$ & $F_{\phi^{+}}=0.73(8)$ \\ 
$0.45(4)$ & $0.36(4)$ & $F_{\phi^{+}}=0.760(16)$ & $F_{\phi^{+}}=0.79(7)$\\ 
$0.17(4)$ & $0.39(5)$ & $F_{\phi^{+}}=0.786(13)$ & $F_{\phi^{+}}=0.80(8)$ \\ 
$0.37(3)$ & $0.45(4)$ & $F_{\phi^{+}}=0.823(8)$ & $F_{\phi^{+}}=0.83(6)$ \\ 
$0.33(4)$ & $0.50(5)$ & $F_{\phi^{+}}=0.743(8)$ & $F_{\phi^{+}}=0.74(6)$ \\ 
$0.34(3)$ & $0.50(10)$ & $F_{\phi^{-}}=0.690(9)$ & $F_{\phi^{-}}=0.71(6)$ \\ 
$0.32(3)$ & $0.53(5)$ & $F_{\phi^{-}}=0.825(9)$ & $F_{\phi^{-}}=0.86(6)$ \\ 
$0.30(3)$ & $0.55(4)$ & $F_{\phi^{+}}=0.561(9)$ & $F_{\phi^{+}}=0.54(7)$ \\ 
$0.21(4)$ & $0.58(4)$ & $F_{\phi^{+}}=0.847(13)$ & $F_{\phi^{+}}=0.87(7)$ \\ 
$0.23(4)$ & $0.65(6)$ & $F_{\phi^{+}}=0.874(13)$ & $F_{\phi^{+}}=0.89(7)$ \\ 
\hline
 \end{tabular}
 \end{center}
\caption{Measured Bell-state fidelities of the states of Figure \ref{fig:TS}, in order of increasing tangle. Fidelities were obtained both with six measurements (equation \ref{eq:fid2}) and sixteen measurements (via quantum state tomography \cite{James}). Note that the Bell-state fidelity is an indicator of entanglement, not a measure.}
 \label{tab:expW}
\end{table}

\section{Process tomography}
\label{sec:process}

As can be seen from the preceding sections, measuring the state is not measuring the gate.  However, in the same manner that a set of output probabilities allows reconstruction of the output state, a set of output states enables reconstruction of the gate operation.  Any quantum operation $\mathcal{E}$ can be written in the \emph{operator-sum representation} for an arbitrary input state $\rho$:
\begin{equation}
\label{eq:op-sum}
\mathcal{E}(\rho) = \sum_k E_k \rho E_k^\dagger
\end{equation}
where $E_k$ are known as \emph{operation elements} or \emph{Kraus operators}, and have the condition that $\sum_j E_j^\dagger E_j\le I$, with equality for trace preserving maps. The set of operation elements $\{E_k\}$  completely describes the effect of the operation including unitary evolution, measurement and decoherence. It should be noted that an operator sum representation is not unique, and there is unitary freedom in it's choice.  With process tomography, we want to experimentally determine some $E_k$ matrices that represent the process.  There are essentially two approaches that can be taken: standard quantum process tomography (SQPT) \cite{97chuang2455,97poyatos390,Nielsen98,0012017,Childs01} in which state tomography is carried out on the result from a set of input states $\{\rho_i\}$; and ancilla-assisted process tomography (AAPT) \cite{0303038,0211133,AAPT2} where state tomography is carried out on the output of the process $(\mathcal{E}\otimes I)(\sigma)$ on a \emph{single} input state $\sigma$  of the original system combined with an ancilla, or auxilliary, system. We present only SQPT here, as both variations achieve the same end, and the initial advantage of SQPT for two qubit gates is the requirement of simpler input states.

We first present the general formulation of the technique and follow this with a much simpler formulation for two qubits where we've chosen a particular basis. Our discussion largely follows references \cite{97chuang2455,MikeIke}. SQPT is performed in the following steps:
\begin{enumerate}
\item We choose a fixed basis of linearly independent input states $\{\tilde{\rho}_i\}$ and experimentally determine the output density operators ${\mathcal{E}(\tilde{\rho}_i)}$ for each using state tomography. These we can express in our basis:
\begin{equation}
\label{eq:op-in-exp}
\mathcal{E}(\tilde{\rho}_i) = \sum_j \lambda_{ij} \tilde{\rho}_j.
\end{equation}
  
\item We also choose a fixed basis $\{\tilde{E}_j\}$ for the operators on the state space. We can express the operation elements in (\ref{eq:op-sum}) in this basis, $E_i = \sum_m a_{im} \tilde{E}_m$, hence,
\begin{equation}
\label{eq:op-in-basis}
\mathcal{E}(\rho) =  \sum_{mn} \chi_{mn} \tilde{E}_m \rho \tilde{E}_n^\dagger,
\end{equation}
where $\chi_{mn}$ is a positive Hermitian matrix, $\chi_{mn}\equiv \sum_i a_{im}a_{in}^*$, that completely describes the process in the chosen basis.
  
\item Now, writing $\tilde{E}_m \tilde{\rho}_i \tilde{E}_n^\dagger = \sum_{j} \beta_{ij}^{mn} \tilde{\rho}_j$ we can express the process for the input states as
\begin{equation}
\label{eq:op-in-theory}
\mathcal{E}(\tilde{\rho}_i) = \sum_{jmn}\chi_{mn}\beta_{ij}^{mn} \tilde{\rho}_j
\end{equation}
  
Comparing equations~(\ref{eq:op-in-exp}) and (\ref{eq:op-in-theory}) and using the fact that the $\tilde{\rho}_i$ are a basis, we get $\lambda_{ij}=\sum_{mn}\beta_{ij}^{mn}\chi_{mn}$. By combining the indices $ij$ and $mn$ (e.g. $\{11,21,31,41,12,\ldots,44\}$), we have the matrix equation $\vec{\lambda} = \beta \vec{\chi}$. Hence by inverting $\beta$ we can find $\vec{\chi}$ in terms of the experimentally determined $\vec{\lambda}$.

\item As mentioned previously, $\chi$ is positive, so by the spectral theorem we can decompose it into a product of some unitary operator $U$ and and a positive diagonal matrix $D$, $\chi=U D U^\dagger$. Hence we can construct the operation elements 
\begin{equation}
\label{eq:op-elem}
E_k =\sqrt{D_{kk}}\sum_{m}U_{mk}\tilde{E}_m
\end{equation}
\end{enumerate}

For particular cases we can choose the basis so that the above calculations are easily performed. Consider a two qubit process.  If the basis for the input states is given by $\rho^{(jk)}_{mn}\equiv \delta_{jm}\delta_{kn}$ where $\rho^{(jk)}$ is a matrix with a 1 on the $j$th row and $k$th column and zeros elsewhere; and we choose as our operator basis $\tilde{A}\otimes\tilde{B}$ where $\tilde{A},\tilde{B}\in \{I,X,Y,Z\}$, then the $16\times 16$ $\chi$ matrix is easily constructed in block form:
\begin{equation}
  \label{eq:out-block}
  \chi =K^T 
\begin{pmatrix}
\mathcal{E}(\rho^{(11)}) & \mathcal{E}(\rho^{(12)}) & \cdots\\
\mathcal{E}(\rho^{(21)}) & \mathcal{E}(\rho^{(22)}) & \cdots\\
\vdots & \vdots & \ddots
\end{pmatrix}
  K 
\end{equation}
where $K$ is a particular matrix constructed from $K=P\Lambda$, with $P = I\otimes [\rho^{(11)} + \rho^{(23)} + \rho^{(32)} + \rho^{(44)}] \otimes I$ and $\Lambda = (Z\otimes I + X\otimes X) \otimes (Z \otimes I + X \otimes X)/4 $. 

The $\rho^{(jk)}$ basis operators for the initial states don't necessarily themselves represent physical states. This is not a problem as we can use the linearity of the operation $\mathcal{E}(\rho)$ to construct them from suitable combinations of other operators. In appendix~\ref{sec:physical-input-basis} we relate $\rho^{(jk)}$ to the $\{\ketbra{H}{H}, \ketbra{V}{V}, \ketbra{D}{D}, \ketbra{R}{R}\}$ operators. In appendix~\ref{sec:example} we give a detailed example of process tomography for a two-qubit gate.

Note that none of these procedures allow for experimental uncertainties and their effect on process measurement --- it is a desirable and urgent matter that this be explored, in much the same manner as has been done for quantum state tomography \cite{James}. 

Having obtained the operator elements, how do we tell how close the process is to the ideal gate we are trying to implement? There are several possible approaches here, from minimisations over input states, to exploiting an isomorphism between processes and states.  We outline some of these below, where we concentrate on fidelity-like measures.

There are several measures that can be constructed from a minimisation over input states --- which can be done numerically since we've already characterised the process. For instance the \emph{gate fidelity} \cite{MikeIke} between a target unitary gate $U$ and the actual process $\mathcal{E}$ can be calculated from
\begin{align}
  \label{eq:gate-fid}
  F_g(U,\mathcal{E}) =& \underset{\ket{\psi}}{\mathrm{min}}\; 
F[U\ketbra{\psi}{\psi}U^\dagger,\mathcal{E}(\ketbra{\psi}{\psi})]\\
 =& \underset{\ket{\psi}}{\mathrm{min}}\; \bra{\psi}U^\dagger
\mathcal{E}(\ketbra{\psi}{\psi})
U\ket{\psi}
\end{align}
where $F$ is the usual state fidelity between two density operators $\rho$ and $\sigma$, $F(\rho, \sigma) = \mathrm{Tr} \sqrt{\sqrt{\sigma} \rho \sqrt{\sigma}}$.  An advantage of this type of measure is that it is based on a well understood state-based measure and there is a clear physical interpretation \cite{Dodd02}.

An alternative that does not involve a minimisation, is the \emph{average gate fidelity} $\bar{F}$ for a process \cite{0205035}. The average gate fidelity is stated for qudits. We can consider our two-qubit system as equivalent to a 4-level qudit so that $\bar{F}$ becomes
\begin{equation} \label{eq:ave-gate-fid}
\bar{F}(U,\mathcal{E})=\frac{1}{5}
+\frac{1}{80}\sum_j \mathrm{Tr}\{U \tilde{U}_j^\dagger
U^\dagger\mathcal{E}(\tilde{U}_j) \}
\end{equation} 
where $\tilde{U}_j$ are a basis of \emph{unitary} operators in the qudit space. For this basis, we can use the same basis we used to expand the operators in the process tomography, i.e. tensor products of Pauli operators. Of course, we will have to calculate the $\mathcal{E}(\tilde{U}_j)$ from suitable linear combinations of physical input states in a manner similar to that given in appendix~\ref{eq:physical basis}.  Note that this measure can be calculated more directly than via the process tomography, since we only need to determine the $\mathcal{E}(\tilde{U}_j)$, and do not need to determine the actual operation elements.

The last approach is to exploit an isomorphism between quantum processes and unnormalised states \cite{72jamiolkowski275,01cirac544}, where we can construct a positive state, $R$, from the action of the process on part of the maximally entangled state $\ket{\Psi_+}=\sum_j \ket{j}\ket{j}$,
\begin{equation}
  \label{eq:process-state}
  R = (\mathcal{E}\otimes I)\ketbra{\Psi^{+}}{\Psi^{+}}
\end{equation}
Once we've obtained $R$ then we can use the usual tools for measuring the distance of two states to compare the two processes, such as, for example, the fidelity \cite{01raginsky11}. 

For all these measures, there are open questions in applying them to experimentally characterising gates. Most notably, how do the errors in the state preparation and state tomography finally manifest themselves in the measure? More subtly, much research needs to be done in identifying appropriate process measures that offer physically significant feedback --- obviously a critical issue for experimentalists trying to improve the implementation of two-qubit gates.

\section{Conclusion}

\begin{table}
\label{ET}
 \begin{center}
 \begin{tabular}{|l|c|} \hline
Measure & Condition \\ \hline
{\em Simple measures} & \\
\phantom{!} Inquisition \phantom{!} &  $1 \geq \mathcal{I} > 0$ \\
\phantom{!} Logical visibility \phantom{!} &  $1 \geq |V_{L}| > 0$ \\
\phantom{!} Equal-weight visibility \phantom{!} &  $1 \geq |V_{E}(n \frac{\pi}{2})| > 0$\\
\phantom{!} Parity \phantom{!} & $1 \geq |\Pi(n \frac{\pi}{2})| > 0$\\
\phantom{!} Bell-state Fidelity \phantom{!} & $1\geq F_{\phi^\pm,\psi^\pm} > \frac{1}{2}$ \\ 
\phantom{!} Entanglement witness \phantom{!} & $-\frac{1}{2} \leq \langle W_{\phi^\pm,\psi^\pm} \rangle < 0$ \\ \hline
{\em State tomography} & \\
\phantom{!} Linear entropy \phantom{!} & \phantom{!}  $0 \leq S_{L} < 1$ \phantom{!} \\
\phantom{!} Tangle \phantom{!} & \phantom{!}  $1\geq T \geq0$ \phantom{!} \\
\phantom{!} \ \ \ \ \ \ \ \ \ \ (for pure states) &  $T=V_{E}(n \frac{\pi}{2})^{2}$ \\
\phantom{!} Bell's inequality \phantom{!} & \phantom{!} $2 \sqrt{2} \geq B > 2$\phantom{!} \\ 
\phantom{!} \ \ \ \ \ \ \ \ \ \ (for pure states) & $B=2\sqrt{1+T^2}$ \\ \hline
{\em Process tomography} & \\
\phantom{!} Gate fidelity \phantom{!} & \phantom{!}  $1\geq F_{g} > 0$ \phantom{!} \\
\phantom{!} Average gate fidelity \phantom{!} & \phantom{!}  $1\geq \bar{F}_{g} > 0$ \phantom{!} \\
\phantom{!} Isomorphism fidelity \phantom{!} & \phantom{!}  $1\geq F_{i} > 0$ \phantom{!} \\ \hline
 \end{tabular}
 \end{center}
\caption{Table of various \textsc{cnot} gate measures and indicators: the value for an ideal \textsc{cnot} gate is to the left of the symbol; the worst limit is to the right. See text for details and caveats.}
 \label{tab:simple}
\end{table}

We have discussed a range of measurements for \textsc{cnot}  gates that are independent of any given architecture. These are summarised in Table \ref{ET}, and range from simple correlation measurements, suitable for developmental stages of a gate, to state measures that allow quick identification of quantum operation, through to full process tomography, which yields the full functional description of the gate. We note that a similar set of arguments can be applied to derive appropriate measures for any two-qubit entangling gate. 

\subsection*{Acknowledgements}
\noindent We wish to acknowledge Joseph~B.~Altepeter, Christopher M. Dawson, Daniel~F.~V.~James, Duncan Mortimer, William~J.~Munro and Michael~A.~Nielsen for invaluable discussions. This work was supported by the Australian Research Council and the US Army Research Office.

\subsection*{Note}
\noindent During preparation of this paper we became aware of an independent experimental study that used a six parameter Entanglement Witness for the $|\psi^{-} \rangle$ state \cite{Barb03}.

\appendix

\section{Entanglement witnesses and Bell-state Fidelity}
\label{sec:ew}

Interestingly, the witnesses of equation (\ref{eq:W2}) can be written as:
\begin{align}
\label{blah} W_{\phi^{\pm}} & =(|\psi^{\mp}\rangle\langle
\psi^{\mp}|)^{T_2}=\frac{1}{2}I\otimes I -
|\phi^{\pm}\rangle\langle \phi^{\pm} | \\
W_{\psi^{\pm}} & = (|\phi^{\mp}\rangle\langle
\phi^{\mp}|)^{T_2}=\frac{1}{2} I\otimes I
-|\psi^{\pm}\rangle\langle \psi^{\pm}|.
\end{align}
If we note that the condition for the detection of an entangled state is $ \langle W_{\psi^{\pm},\phi^{\mp}}\rangle = \mathrm{Tr}(W_{\psi^{\pm},\phi^{\pm}}\hat{\rho})<0$ then we see:
\begin{align}
\mathrm{Tr}(W_{\psi,\phi}\hat{\rho}) & =
\frac{1}{2}-\mathrm{Tr}(\hat{\rho}|\psi,\phi\rangle\langle\psi,\phi|)\\
& = \frac{1}{2} - F_{\psi^{\pm},\phi^{\pm}},
\end{align}
so if $\langle W_{\psi,\phi}\rangle <0$ we see that this is equivalent to $F_{\psi,\phi}>\frac{1}{2}$, which is simply the Bell-state Fidelity entanglement indicator. Generally, we can derive an optimal class of entanglement witnesses for any two-qubit maximally entangled state in a similar fashion. We note that all of the maximally entangled states on two qubits can be generated from one of the Bell-states and the application of a single local unitary gate, for instance, $U \otimes I |\phi^{+}
\rangle$. Then from \cite{02guhne062305} we know that the optimal witness for the state $(1-p)I\otimes I + p|\phi^{+}\rangle\langle \phi^{+}|$ is $(|\psi^{-}\rangle\langle \psi^{-}|)^{T_2}$. This is
found as $|\psi^{-}\rangle$ is the eigenvector corresponding to the minimum eigenvalue of the matrix $(|\phi^{+}\rangle\langle \phi^{+}|)^{T_2}$. That is:
\begin{equation}
(|\phi^{+}\rangle\langle \phi^{+}|)^{T_2}|\psi^{-}\rangle =
w|\psi^{-}\rangle.
\end{equation}
If we multiply from the left by $U\otimes I$, we get:
\begin{equation}
U\otimes I(|\phi^{+}\rangle\langle
\phi^{+}|)^{T_2}|\psi^{-}\rangle = wU\otimes I|w\psi^{-}\rangle.
\end{equation}
Then if we insert an identity and note that $U\otimes I$ commutes with the partial transpose operation we find:
\begin{equation}
(U\otimes I|\phi^{+}\rangle\langle \phi^{+}|U^{\dagger}\otimes
I)^{T_2}U\otimes I|\psi^{-}\rangle = wU\otimes I|\psi^{-}\rangle,
\end{equation}
hence the optimal witness for an arbitrary Werner state, $(1-p)I\otimes I + pU\otimes I|\phi^{+}\rangle\langle \phi^{+}|U^{\dagger}\otimes I$, is $(U\otimes I|\psi^{-}\rangle\langle \psi^{-}| U^{\dagger}\otimes I)^{T_2}$. Then equation \ref{blah} tells us that this witness can be written as $\frac{1}{2} I\otimes I -U\otimes I |\phi^{+}\rangle\langle \phi^{+}|U^{\dagger}\otimes I$.

Hence we see given some Werner state of the form $pI\otimes I+(1-p) |\theta\rangle\langle \theta|$, the optimal witness the state as defined in \cite{02guhne062305}, is simply $W=\frac{1}{2}I\otimes I -|\theta\rangle\langle\theta|$ where $|\theta\rangle$ is some maximally entangled state. Thus the value of an optimal Werner witness measurement $\langle W \rangle$ is always linearly related to the fidelity between a maximally entangled state and the state being analyzed. This, allows us to interpret this witness as the witness for which the fidelity between the output state and some maximally entangled state is greater than a half.

In general, {\em any} projective measurement that projects on to an entangled state can be used as an entanglement witness in a similar fashion. It is relatively easy to prove that given some entangled state $|\psi\rangle$, with maximum Schmidt coefficient, $\lambda_{max}$, then $\mathrm{Tr}(\hat{\rho}_{sep}|\psi\rangle\langle\psi |)\leq \lambda_{max}^{2}$ for all separable states, $\hat{\rho}_{sep}$. Thus an entanglement witness $W= \lambda_{\max}^{2} I\otimes I - |\psi\rangle\langle\psi|$ can always be constructed from an arbitrary entangled state $|\psi\rangle$.

\section{Physical input basis for process tomography} 
\label{sec:physical-input-basis}
\newcommand{\imag}{i}

The basis operators $\hat{\rho}^{(jk)}$ we used for the process tomography, were matrices with a single 1 in the $j$th row and $k$th column.  We can relate this to some other basis such as $\hat{\rho}^{(\alpha\beta)}=\hat{\rho}^{(\alpha)}\otimes\hat{\rho}^{(\beta)}$ where $\alpha,\beta\in
\{H,V,D,R\}$. Clearly, it is trivial to write the $\hat{\rho}^{(\alpha\beta)}$ operators in terms of combinations of $\hat{\rho}^{(jk)}$. If we write this mapping as a matrix $\vec{\rho}^{\ (\alpha\beta)} = M\vec{\rho}^{\ (jk)}$ we can now simply invert the matrix: $\vec{\rho}^{\ (jk)} = M^{-1}\vec{\rho}^{\ (\alpha\beta)}$, i.e., where $a=(1+i)/2$,
\begin{widetext}
\begin{equation*}
  \label{eq:physical basis}
  \left(\begin{array}{c}\rho^{(11)}\\\rho^{(12)}\\\rho^{(13)}\\\rho^{(14)}\\
\rho^{(21)}\\\rho^{(22)}\\\rho^{(23)}\\\rho^{(24)}\\
\rho^{(31)}\\\rho^{(32)}\\\rho^{(33)}\\\rho^{(34)}\\
\rho^{(41)}\\\rho^{(42)}\\\rho^{(43)}\\\rho^{(44)}
      \end{array}\right)
   = 
  \left(\begin{array}{cccccccccccccccc}
    1 & 0 & 0 & 0 & 0 & 0 & 0 & 0 & 0 & 0 & 0 & 0 & 0 & 0 & 0 & 0 \\
    -a & -a & 1 & \imag & 0 & 0 & 0 & 0 & 0 & 0 & 0 & 0 & 0 & 0 & 0 & 0 \\
    -a & 0 & 0 & 0 & -a & 0 & 0 & 0 & 1 & 0 & 0 & 0 & \imag  & 0 & 0 & 0 \\
    \frac{\imag }{2} & \frac{\imag }{2} & -a & {a^*} & \frac{\imag
    }{2} & \frac{\imag }{2} & -a & {a^*} & -a & -a & 1 & \imag
    & {a^*} & {a^*} & \imag  & -1 \\
    -{a^*} & -{a^*} & 1 & -\imag
    & 0 & 0 & 0 & 0 & 0 & 0 & 0 & 0 & 0 & 0 & 0 & 0 \\
    0 & 1 & 0 & 0 & 0 & 0 & 0 & 0 & 0 &
    0 & 0 & 0 & 0 & 0 & 0 & 0 \\
    \frac{1}{2} & \frac{1}{2} & -a & -{a^*} & \frac{1}{2} &
    \frac{1}{2} & -a & -{a^*} & -{a^*} & -{a^*} & 1 & -\imag & -a & -a
    & \imag
    & 1 \\ 0 & -a & 0 & 0 & 0 & -a & 0 & 0 & 0 & 1 & 0 & 0 & 0 & \imag  & 0 & 0 \\
    -{a^*} & 0 & 0 & 0 & -{a^*} & 0 & 0 & 0 & 1 & 0 & 0 & 0 & -\imag &
    0 & 0 & 0 \\ \frac{1} {2} & \frac{1}{2} & -{a^*} & -a &
    \frac{1}{2} & \frac{1}{2} & -{a^*} & -a & -a & -a & 1 & \imag &
    -{a^*} & -{a^*} & -\imag & 1 \\ 0 & 0 & 0 & 0 & 1 & 0 & 0 & 0 & 0
    & 0 & 0 & 0 & 0 & 0 & 0 & 0 \\ 0 & 0 & 0 & 0 & -a & -a & 1 & \imag
    & 0 & 0 & 0 & 0 & 0 & 0 & 0 & 0 \\ \frac{-\imag }{2} & \frac
    {-\imag }{2} & -{a^*} & a & \frac{-\imag }{2} & \frac{-\imag }{2}
    & -{a^*} & a & -{a^*} & -{a^*} & 1 & -\imag & a & a & -\imag & -1
    \\ 0 & -{a^*} & 0 & 0 & 0 & -{a^*} & 0 & 0 & 0 & 1 & 0 & 0 & 0 &
    -\imag & 0 & 0 \\ 0 & 0 & 0 & 0 & -{a^*} & -{a^*} & 1 & -\imag & 0
    & 0 & 0 & 0 & 0 & 0 & 0 & 0 \\ 0 & 0 & 0 & 0 & 0 & 1 & 0 & 0 & 0 &
    0 & 0 & 0 & 0 & 0 & 0 & 0
  \end{array}\right)
  \left(\begin{array}{c}\rho^{(HH)}\\\rho^{(HV)}\\\rho^{(HD)}\\\rho^{(HR)}\\
\rho^{(VH)}\\\rho^{(VV)}\\\rho^{(VD)}\\\rho^{(VR)}\\
\rho^{(DH)}\\\rho^{(DV)}\\\rho^{(DD)}\\\rho^{(DR)}\\
\rho^{(RH)}\\\rho^{(RV)}\\\rho^{(RD)}\\\rho^{(RR)}
      \end{array}\right).
\end{equation*}
\end{widetext}

\section{Process tomography of a 2-qubit gate --- an example} \label{sec:example} 
First we need a basis of input states. For each qubit, use the following input states
\begin{equation}
  \begin{split}
    \hat{\rho}^{(H)} = \left(\begin{array}{cc} 1&0\\0&0\end{array}\right)
    \quad \hat{\rho}^{(V)} = \left(\begin{array}{cc}
        0&0\\0&1\end{array}\right) \\
    \hat{\rho}^{(D)} =
    \frac{1}{2}\left(\begin{array}{cc} 1&1\\1&1\end{array}\right)
    \quad \hat{\rho}^{(R)} = \frac{1}{2}\left(\begin{array}{cc} 1&-i\\ 
        i&1\end{array}\right)
\end{split}
\end{equation}
so that the basis for the two qubits comprises of 16 two-qubit states given  by all the tensor products of the single qubit states $\{\hat{\rho}^{(\alpha\beta)}\}=\{\hat{\rho}^{(\alpha)}\otimes\hat{\rho}^{(\beta)}\}$, $\alpha,\beta \; \in \; \{H,V,D,R\}$. For example,
\begin{equation}
\hat{\rho}^{(DR)} = \frac{1}{4} 
\left(\begin{array}{cccc} 1&-i&1&-i \\ 
    i&1&i&1\\ 
    1&-i&1&-i \\ 
    i&1&i&1\end{array}\right)
\end{equation}
Using $\{\hat{\rho}^{(\alpha\beta)}\}$ as input states and performing state tomography on each one we obtain a set of output matrices $\{\mathcal{E}(\hat{\rho}^{(\alpha\beta)})\}$.

Using the transformation in appendix~\ref{sec:physical-input-basis} we can easily calculate the 16 $\{\mathcal{E}(\hat{\rho}^{(jk)})\}$ matrices we need to construct the $\chi$ matrix (\ref{eq:out-block}). For example, with $a=(1+i)/2$,
\begin{equation}
  \begin{split}
  \label{eq:eg.HV->(jk)}
  \mathcal{E}(\hat{\rho}^{(13)}) =& -a \mathcal{E}(\hat{\rho}^{(HH)})
  -a \mathcal{E}(\hat{\rho}^{(VH)})\\
  &+ \mathcal{E}(\hat{\rho}^{(DH)})
 + i  \mathcal{E}(\hat{\rho}^{(RH)}). 
\end{split}
\end{equation}
For the purposes of the illustration let us consider the following process, which applies a \textsc{CNOT} gate with probability $p$ else does nothing,
\begin{equation} 
\label{eq:bit-flipped-cnot} 
  \mathcal{E}(\hat{\rho}) = p\textsc{CNOT}\hat{\rho}\textsc{CNOT}+(1-p)\hat{\rho},
\end{equation} 
and imagine that we obtain the output state $\mathcal{E}(\hat{\rho})$ by state tomography. Constructing the $\chi$ matrix in block-form as given in (\ref{eq:out-block}) we arrive at,
\begin{equation*}
\chi = 
  \frac{1}{4}\left(\begin{array}{ccccccccc}
      4-3p & p & 0 &   & 0 & p & -p & 0 & 0 \\
      p & p & 0 & \cdots & 0 & p & -p & 0 & 0 \\
      0 & 0 & 0  &  & 0 & 0 & 0 & 0 & 0 \\
        & \vdots &   &  &  &  & \vdots &  &  \\
      0 & 0 & 0  &   & 0 & 0 & 0 & 0 & 0 \\
      p & p & 0 &   & 0 & p & -p & 0 & 0 \\
      -p & -p & 0 & \cdots  & 0 & -p & p & 0 & 0 \\
      0 & 0 & 0 &  & 0  & 0  & 0 & 0 & 0 \\
      0 & 0 & 0 &  & 0  & 0  & 0 & 0 & 0 
    \end{array}\right).
\end{equation*}
It is fairly easy to ``read'' the $\chi$ matrix directly --- consider the extreme cases of $p=0$ and $p=1$. For $p=0$ only the top right element is non-zero and this corresponds to $I\otimes I$, i.e. no transformation is applied. For $p=1$, writing out the element from the matrix according to equation~(\ref{eq:op-in-basis}) and factorising,
\begin{align}
  \label{eq:eg-cnot-decomp}
  \mathcal{E}(\hat{\rho}) &= U\hat{\rho} U^\dagger\\
  U &= \frac{1}{2}(I\otimes I+I\otimes X+Z\otimes I-Z\otimes X),
\end{align}
and it is readily verified that $U=\textsc{CNOT}$.


\begin{thebibliography}{99}
\bibitem{Comb} For the target in the logical 1 state, the fringe visibility is simply $V_{L}=(P_{odd}-P_{even})/(P_{odd}+P_{even})$.
\bibitem{CircleNote} Note that on the qubit characteristic sphere the equal-weight superpositions form a great circle, perpendicular to the line joining the logical states. For example, they lie on the equator of the Bloch sphere, and can be accessed with $\pi/2$ pulses.
\bibitem{PTfoot} The partial transpose of $\hat{\rho}$, with respect to an orthonormal basis $\{\ket{b}\}$ of qubit two, is given by $\hat{\rho}^{T_2}=\sum_{bb'} \ketbra{b'}{b}\hat{\rho}\ketbra{b'}{b}$.
\end{thebibliography}
\end{document}